\documentclass[12 pt]{elsarticle}
\usepackage{lineno,hyperref}
\modulolinenumbers[5]
\usepackage{amsmath}
\usepackage[utf8]{inputenc}
\usepackage[english]{babel}
\usepackage{mathtools}
\usepackage{amsthm}
\usepackage{amssymb}
\usepackage{multirow}
\usepackage{relsize}

\newcommand{\bt}{\begin{table}[!h]}
\newcommand{\et}{\end{table}}
\newcommand{\be}{\begin{equation}}
\newcommand{\ee}{\end{equation}}

\makeatletter
\def\ps@pprintTitle{%
\let\@oddhead\@empty
\let\@evenhead\@empty
\def\@oddfoot{\centerline{\thepage}}%
\let\@evenfoot\@oddfoot}
\makeatother
\begin{document}
\begin{frontmatter}
\title{ A Stress-Strength Reliability Model using Exponential-Gamma$(3,\lambda)$ Distribution}
\author[1]{ Beenu Thomas }
\author[2]{ V. M. Chacko}
%\affil[1,3]{Department of Statistics, St.Thomas' College(Autonomous), Thrissur, Kerala, India}
%\affil[2]{Department of Mathematics and Statistics, IIT Kanpur}

%\address{Department of Statistics}

\author[mysecondaryaddress]{\\St. Thomas' College (Autonomous)Thrissur, Kerala, India}

\begin{abstract}
One of the important problem in reliability analysis is computation of stress-strength reliability. But it is impractical to compute it in certain situations.
So the estimation stay as an alternative solution to get an approximate value of the reliability. There are research papers which deals with stress-strength reliability analysis using statistical distributions. 
In this paper, a stress-strength reliability model for exponential-gamma$(3,\lambda)$ distribution is introduced. The maximum likelihood estimator (MLE) for the model parameters is derived. Asymptotic distribution and confidence interval for the maximum likelihood estimates of stress-strength reliability, $R=P(X>Y)$, are given. The numerical illustration is performed using Monte Carlo simulations. The results are analyzed with real data analysis.
\end{abstract}

\begin{keyword}
Stress-strength reliability\sep Exponential-Gamma distribution \sep Statistical inference \sep Maximum likelihood estimators

\end{keyword}

\end{frontmatter}

\section{Introduction}
The strength of a unit can be treated as a random variable. Due to uncertainty, the unit's stress should also be treated as a random variable in its operating context. Let X represent a unit's strength and Y represent the random stress that the operational environment imposes on the unit. $R = P(X>Y)$ is the formula for defining a unit's stress-strength reliability (R). A substantial amount of literature exists regarding the problems associated with the estimation of reliability for single-component stress-strength models.
\vskip.1cm
The Exponential-Gamma$(3,\lambda)$ distribution has been proposed by Thomas and Chacko (2020) in conjunction with a bathtub-shaped failure rate function. In this paper, we denote the Exponential-Gamma Distribution$(3,\lambda)$ by $EGD(3,\lambda)$. Specifically, $EGD(3,\lambda)$ has the PDF
\begin{equation}\label{pdfegd}
  f(x)=\frac{\lambda^2}{1+\lambda} (1+\frac{\lambda}{2}x^2) e^{-\lambda x}, x > 0, \lambda > 0.
\end{equation}
It should be noted that $EGD(3,\lambda)$ is a mixture of exponential distribution with a scale parameter of $\lambda$ and gamma distribution with a shape parameter of 3 and a scale parameter of $\lambda$ with mixing proportion $\frac{\lambda}{1+\lambda}$.
\vskip.1cm
Consider two independent random variables X and Y from the $EGD(3,\lambda)$ with different parameters $\lambda_1$ and $\lambda_2$. This paper focuses on the estimation of the parameter $R=P(X<Y)$. Typically, the problem of estimating R arises when dealing with the reliability of a component of strength X subject to a load or stress Y. The component will fail if $X<Y$. As a result, R can be viewed as a measure of component reliability.
\vskip.1cm There has been a long history of stress-strength (SS) reliability, beginning with the pioneering work of Birnbaum (1956) and Birnbaum and McCarty (1958). Church and Harris (1970) are credited with introducing the term stress-strength. Kotz et al. (2003) provide an excellent overview of the various stress-strength models up to 2001. Several publications on the stress-strength model have been published recently, including Gupta and Brown (2001), Raqab and Kundu (2005), Kundu and Gupta (2005, 2006), Krishnamoorthy et al. (2007), Kundu and Raqab (2009), Sharma et al.(2015) and their references.
\vskip.1cm
Based on a bivariate Pareto model, Hanagal (1997) calculated the maximum likelihood estimate (MLE) of the SS parameter R. The estimation of R for three-parameter generalized exponential distributions was investigated by Raqaab et al. (2008). Balakrishnan and Lai (2009) estimated R in models with correlated stress and strength. A study by Al-Mutairi et al. (2013) examined R-estimates for Lindley distributions. The estimation of reliability $R = P(Y< X)$ where X and Y are independent random variables that follow the Kumaraswamy distribution with varying parameters was discussed by Nadar et al. (2014). Ghitany et al. (2015) discussed the reliability of SS systems based on power Lindley distributions. For a transmuted Rayleigh distribution, Dey et al. (2017) calculated the SS reliability parameter R. Deepthi and Chacko (2020) discussed the single-component SS reliability procedure and multi-component SS reliability estimation for the three-parameter generalized Lindley distribution. Varghese and Chacko (2022) examined the reliability of the SS model for the Akash distribution.
\vskip.1cm
Our goal in this article is to estimate the reliability of SS reliability when both stress and strength follow EGD with different parameters $\lambda_1$ and $\lambda_2$. Section 2 considers the SS reliability of $EGD(3, \lambda)$. In section 3, the MLE of R, asymptotic distribution, and CI for the MLE of R are obtained. Section 4 illustrates the extensive simulation study. In section 5, the results are presented with real data analysis. In the final section, conclusions are given.

\section{Stress-Strength Reliability of $EGD(3, \lambda)$ Distribution}
SS reliability is estimated using the EGD distribution in this section.
The SS reliability for the independent random variables X and Y is given by
\begin{equation*}
  R=\int_{-\infty}^\infty f_X(x) \:F_Y(x)\: dx,
\end{equation*}
where $f_X(x)$ and $F_Y(x)$ are the marginal PDF of X and marginal CDF of Y, respectively.
\vskip.1cm We next consider the EGD distribution. Consider X and Y as independent random variables complying with the EGD distribution with parameters $\lambda_1$ and $\lambda_2$, respectively. Thus, we denote $X \sim EGD(3, \lambda_1)$ and $Y \sim EGD(3, \lambda_2)$. Then, SS reliability is

\begin{equation*}
 R=\int_0^\infty \frac{\lambda_1^2}{(1+\lambda_1)}\bigg(1+\frac{\lambda_1}{2}x^2\bigg) e^{-\lambda_1 x}\bigg[1-\frac{(\lambda_2(x(\lambda_2 x+2)+2)+2)e^{-\lambda_2 x}}{2(1+\lambda_2)}\bigg] dx
\end{equation*}
\begin{small}
\begin{equation*}
    =\frac{\lambda_1^2}{2(1+\lambda_1)(1+\lambda_2)}\int_0^\infty \bigg(1+\frac{\lambda_1 x^2}{2}\bigg) e^{-\lambda_1 x}[2(1+\lambda_2)-(\lambda_2^2x^2+2x\lambda_2+2\lambda_2+2)e^{-\lambda_2 x}]dx
\end{equation*}
\end{small}

\begin{align*}
   =\frac{\lambda_1^2}{(1+\lambda_1)}\int_0^\infty \bigg(1+\frac{\lambda_1}{2}x^2\bigg) e^{-\lambda_1 x} dx&-\frac{\lambda_1^2}{2(1+\lambda_1)(1+\lambda_2)}\int_0^\infty \bigg(1+\frac{\lambda_1}{2}x^2\bigg)\\
   & e^{-(\lambda_1 + \lambda_2 )x}(\lambda_2^2x^2+2x\lambda_2+2\lambda_2+2)dx
   \end{align*}
\begin{align*}
   =1-&\biggr[\frac{\lambda_1^2(\lambda_2^2+\lambda_1\lambda_2+\lambda_1)}{(1+\lambda_1)(1+\lambda_2)(\lambda_1+\lambda_2)^3}+\frac{\lambda_1^2\lambda_2}{(1+\lambda_1)(1+\lambda_2)(\lambda_1+\lambda_2)^2}+\frac{\lambda_1^2}{(1+\lambda_1)(\lambda_1+\lambda_2)}\\
   &+\frac{6\lambda_1^3\lambda_2^2}{(1+\lambda_1)(1+\lambda_2)(\lambda_1+\lambda_2)^5}+\frac{3\lambda_1^3\lambda_2}{(1+\lambda_1)(1+\lambda_2)(\lambda_1+\lambda_2)^4}\biggr]
\end{align*}
\begin{align}\label{egdssvalue}
   =&\frac{\lambda_2(10\lambda_1^2\lambda_2^2+5\lambda_1\lambda_2^3+\lambda_2^4+12\lambda_1^2\lambda_2^3+6\lambda_1\lambda_2^4)}{(1+\lambda_1)(1+\lambda_2)(\lambda_1+\lambda_2)^5}+\\ \nonumber
   &\frac{\lambda_2(+3\lambda_1^4\lambda_2+10\lambda_1^3\lambda_2^2+\lambda_2^5+\lambda_1^5\lambda_2+4\lambda_1^4\lambda_2^2+6\lambda_1^3\lambda_2^3+4\lambda_1^2\lambda_2^4+\lambda_1\lambda_2^5)}{(1+\lambda_1)(1+\lambda_2)(\lambda_1+\lambda_2)^5}
\end{align}

\section{Maximum Likelihood Estimate of R}
Let us suppose that $x_1, x_2,\dots, x_n$ is a random sample of size \textit{n} from $EGD(3, \lambda_1)$ and $y_1, y_2,\dots, y_m$ is a random sample of size \textit{m} from $EGD(3, \lambda_2)$.
 The likelihood function is
 \begin{equation}\label{likeliegdss}
   L =\frac{\lambda_1^{2n}}{(1+\lambda_1)^n} e^{-\lambda_1\sum_{i=1}^n x_i} \frac{\lambda_2^{2m}}{(1+\lambda_2)^m}e^{-\lambda_2 \sum_{j=1}^n y_j}\prod_{i=1}^n\big(1+\frac{\lambda_1}{2}x_i^2\big)\prod_{j=1}^m\big(1+\frac{\lambda_2}{2}y_j^2\big)
 \end{equation}
The log-likelihood associated with the above equation is given by
\begin{align}\label{loglssegd}
  \log L = 2n\log \lambda_1 - n\log(1+\lambda_1) + 2m \log \lambda_2 - m \log(1+\lambda_2)-\lambda_1\sum_{i=1}^n x_i-\lambda_2\sum_{j=1}^{m}y_j\\ \nonumber
  +\sum_{i=1}^n\log(1+\frac{\lambda_1}{2}x_i^2)+\sum_{j=1}^m\log(1+\frac{\lambda_2}{2}y_j^2)
\end{align}
The first derivative of Eq.\ref{loglssegd} with respect to the unknown parameters $\lambda_1$ and $\lambda_2$ are respectively given by
\begin{equation*}\label{norm1egdss}
  \frac{\partial\log L}{\partial \lambda_1}=\frac{2n}{\lambda_1}-\frac{n}{1+\lambda_1}-\sum_{i=1}^nx_i+\sum_{i=1}^n\frac{x_i^2}{2(1+\frac{\lambda_1}{2}x_i^2)}
\end{equation*}

\begin{equation*}\label{norm2egd}
\frac{\partial\log L}{\partial \lambda_2}=\frac{2m}{\lambda_2}-\frac{m}{1+\lambda_2}-\sum_{j=1}^m+\sum_{j=1}^m\frac{y_j^2}{2(1+\frac{\lambda_2}{2}y_j^2)}.
\end{equation*}
The second derivative of Eq.\ref{loglssegd} with respect to the unknown parameters $\lambda_1$ and $\lambda_2$ are respectively given by
\begin{equation*}
  \frac{\partial^2\log L}{\partial\lambda_1^2}=\frac{n}{(1+\lambda_1)^2}-\frac{2n}{\lambda_1^2}-\sum_{i=1}^n\frac{x_i^4}{4(1+\frac{\lambda_1}{2}x_i^2)^2}
\end{equation*}
\begin{equation*}
  \frac{\partial^2\log L}{\partial\lambda_2^2}=\frac{m}{(1+\lambda_2)^2}-\frac{2m}{\lambda_2^2}-\sum_{j=1}^m\frac{y_j^4}{4(1+\frac{\lambda_2}{2}y_j^2)^2}
\end{equation*}
According to Eq.\ref{egdssvalue}, MLE of SS reliability, $\hat{R}_{ML}$, can be calculated as follows:
%\begin{tabular}
%\renewcommand*{\arraystretch}{1}

\begin{equation}\label{mleegdssr}
\mathsmaller{
  \hat{R}_{ML} =\frac{\lambda_2(10\lambda_1^2\lambda_2^2+5\lambda_1\lambda_2^3+\lambda_2^4+12\lambda_1^2\lambda_2^3+6\lambda_1\lambda_2^4+3\lambda_1^4\lambda_2+10\lambda_1^3\lambda_2^2+\lambda_2^5+\lambda_1^5\lambda_2+4\lambda_1^4\lambda_2^2+6\lambda_1^3\lambda_2^3+4\lambda_1^2\lambda_2^4+\lambda_1\lambda_2^5)}{(1+\lambda_1)(1+\lambda_2)(\lambda_1+\lambda_2)^5}
}\end{equation}
%\end{tabular}

\subsection*{\textbf{Asymptotic Distribution and Confidence Intervals}}
The asymptotic distribution and confidence interval (CI) for the MLE of R are presented in this section. Let us represent the Fisher information matrix of $\lambda= (\lambda_1, \lambda_2)$ as $I(\lambda)$ in order to obtain the asymptotic variance of the MLE of R, $\hat{R}_{ML}$, in Eq.\ref{mleegdssr}.
$$
I(\lambda)=E
\left[ {\begin{array}{cc}
	-\frac{\partial^{2}\log L}{\partial\lambda_1^{2}} & -\frac{\partial^{2}\log L}{\partial\lambda_1\partial\lambda_2} \\[8pt]
	-\frac{\partial^{2}\log L}{\partial\lambda_2\partial\lambda_1} & -\frac{\partial^{2}\log L}{\partial\lambda_2^{2}} \\
	\end{array} } \right].
$$
\\
We further define R to establish its asymptotic normality as
\begin{equation*}
  d(\lambda) = \bigg(\frac{\partial R}{\partial \lambda_1}, \frac{\partial R}{\partial \lambda_2}\bigg)^\prime = (d_1, d_2)^\prime
\end{equation*}
where
\begin{equation*}
\mathsmaller{
  \frac{\partial R}{\partial \lambda_1}=-\frac{\lambda_1\lambda_2^2(\lambda_1^5+(4\lambda_2+6)\lambda_1^4+(+\lambda_2^2+20\lambda_2+3)\lambda_1^3+(4\lambda_2^3+24\lambda_2^2+48\lambda_2)\lambda_1^2+(\lambda_2^4+12\lambda_2^3+21\lambda_2^2+30\lambda_2)\lambda_1+2\lambda_2^4+6\lambda_2^3)}{(1+\lambda_1)^2(1+\lambda_2)(\lambda_1+\lambda_2)^6}
}\end{equation*}

\begin{equation*}
\mathsmaller{
  \frac{\partial R}{\partial \lambda_2}=\frac{\lambda_1^2\lambda_2(\lambda_2^5+2(2\lambda_1+3)\lambda_2^4+(6\lambda_1^2+20\lambda_1)\lambda_2^3+4\lambda_1(\lambda_1^2+6\lambda_1+12)\lambda_2^2+\lambda_1(\lambda_1^3+12\lambda_1^2+21\lambda_1+30)\lambda_2+2\lambda_1^4+6\lambda_1^3}{(1+\lambda_1)^2(1+\lambda_2)(\lambda_1+\lambda_2)^6}
}\end{equation*}
As a result, we can find the asymptotic distribution of $\hat{R}_{ML}$ as
\begin{equation*}
  \sqrt{n+m} (\hat{R}_{ML}-R)\rightarrow^d N (0, d^\prime(\lambda)\: I^{-1}(\lambda)\: d(\lambda))
\end{equation*}
We obtain the asymptotic variance of $\hat{R}_{ML}$ as follows:
\begin{align*}
  AV (\hat{R}_{ML})&= \frac{1}{n+m}d^\prime(\lambda)\: I^{-1}(\lambda)\: d(\lambda)\\
  &=V(\hat{\lambda_1}) d_1^2+V(\hat{\lambda_1}) d_2^2+2 d_1 d_2 (\hat{\lambda_1}\hat{\lambda_1}).
\end{align*}
 Asymptotic $100(1-\omega)\%$ CI for R can be obtained as
 \begin{equation*}
   \hat{R}_{ML} \pm Z_{\omega/2}\sqrt{AV(\hat{R}_{ML})}
 \end{equation*}
where $Z_{\omega/2}$ is the upper $\omega/2$ quantile of the standard normal distribution.

\section{Simulation Study}
This section presents some results related to the performance of estimators in R using the Newton-Raphson method. Using independent $EGD(3, \lambda_1)$ and $EGD(3, \lambda_2)$ distributions, we have generated 1000 samples for this purpose. The parameter values, $(\lambda_1, \lambda_2)$, examined in this study were: (0.5, 1.5), (1, 1.5), and (1, 0.5), and various sample sizes (n, m): (10,10), (15,15), (25,25), (30,30), (50,50), and (75,75). According to these parameter values, R values are 0.8391, 0.6405, and 0.2551, respectively.

\vskip.1cm Tables \ref{simegdss1}- \ref{simegdss3} provide estimates of R based on the MLE method along with average biases, mean square errors (MSEs), and 95\% CIs. Simulated results indicate that biases and MSEs decrease with increasing sample size (n,m).

\begin{table}[!ht]
    \caption{\textbf{MLE, average(Avg) bias, and MSEs of different estimators of R when $\lambda_1=0.5$ and $\lambda_2=1.5$.}}\label{simegdss1}
    \centering
    \begin{tabular}{llllr}
    \hline
        \textbf{(n,m)} & \textbf{Estimates} & \textbf{Avg Bias} & \textbf{MSEs} & \textbf{95\% CI}\\
         \hline
        \rule{0pt}{10pt}\multirow{2}{*}{(10,10)} & $\hat{\lambda_1}$=0.52881 & 0.02881 & 0.01272 &(0.00714, 1.67700)\\
            & $\hat{\lambda_2}$=1.62901 & 0.12901 & 0.19516  \\
            \hline
     \rule{0pt}{10pt}\multirow{2}{*}{(15,15)} & $\hat{\lambda_1}$=0.51921 & 0.019208 & 0.00808 &(-0.16250, 1.84527)\\
            & $\hat{\lambda_2}$=1.58803 & 0.08803 & 0.11591 \\
            \hline
      \rule{0pt}{10pt}\multirow{2}{*}{(25,25)} & $\hat{\lambda_1}$=0.51427 & 0.01427 & 0.00453 &(0.29010, 1.38693)\\
            & $\hat{\lambda_2}$=1.54604 & 0.04604 & 0.05462 \\
            \hline
       \rule{0pt}{10pt}\multirow{2}{*}{(30,30)} & $\hat{\lambda_1}$=0.50998 & 0.00998 & 0.00389 &(0.08218, 1.59616)\\
            & $\hat{\lambda_2}$=1.53623 & 0.03623 & 0.04227 \\
            \hline
       \rule{0pt}{10pt}\multirow{2}{*}{(50,50)} & $\hat{\lambda_1}$=0.50531 & 0.00531 & 0.00221 &(0.41299, 1.26788)\\
            & $\hat{\lambda_2}$=1.53003 & 0.03003 & 0.02474 \\
            \hline
      \rule{0pt}{10pt}\multirow{2}{*}{(75,75)} & $\hat{\lambda_1}$=0.50529 & 0.00529 & 0.00150 &(0.42865, 1.24868)\\
            & $\hat{\lambda_2}$=1.51549 & 0.01549 & 0.01516 \\
            \hline
    \end{tabular}
\end{table}
\begin{table}[!ht]
    \caption{\textbf{MLE, average(Avg) bias, and MSEs of different estimators of R when $\lambda_1=1$ and $\lambda_2=1.5$.}}\label{simegdss2}
    \centering
    \begin{tabular}{llllr}
    \hline
        \textbf{(n,m)} & \textbf{Estimates} & \textbf{Avg Bias} & \textbf{MSEs} & \textbf{95\% CI}\\
         \hline
        \rule{0pt}{10pt}\multirow{2}{*}{(10,10)} & $\hat{\lambda_1}$=1.07005 & 0.07005 & 0.07308 &(-0.95120, 2.23538)\\
            & $\hat{\lambda_2}$=1.62033 & 0.12033 & 0.17647 \\
            \hline
     \rule{0pt}{10pt}\multirow{2}{*}{(15,15)} & $\hat{\lambda_1}$=1.03367 & 0.03367 & 0.03485 &(-0.30554, 1.59834)\\
            & $\hat{\lambda_2}$=1.58321 & 0.08321 & 0.11195 \\
            \hline
      \rule{0pt}{10pt}\multirow{2}{*}{(25,25)} & $\hat{\lambda_1}$=1.02501 & 0.02501 & 0.02182 &(-0.36708, 1.65794)\\
            & $\hat{\lambda_2}$=1.56419 & 0.06418 & 0.06247 \\
            \hline
       \rule{0pt}{10pt}\multirow{2}{*}{(30,30)} & $\hat{\lambda_1}$=1.01862 & 0.01862 & 0.01575 &(-0.05805, 1.34368)\\
            & $\hat{\lambda_2}$=1.54095 & 0.04095 & 0.04532 \\
            \hline
       \rule{0pt}{10pt}\multirow{2}{*}{(50,50)} & $\hat{\lambda_1}$=1.01643 & 0.01643 & 0.01089 &(0.18678, 1.09807)\\
            & $\hat{\lambda_2}$=1.53552 & 0.03552 & 0.02541 \\
            \hline
      \rule{0pt}{10pt}\multirow{2}{*}{(75,75)} & $\hat{\lambda_1}$=1.01383 & 0.01383 & 0.00614 &(0.17796, 1.10334)\\
            & $\hat{\lambda_2}$=1.52276 & 0.02276 & 0.01679 \\
            \hline
    \end{tabular}
\end{table}
\begin{table}[!ht]
    \caption{\textbf{MLE, average(Avg) bias, and MSEs of different estimators of R when $\lambda_1=1$ and $\lambda_2=0.5$.}}\label{simegdss3}
    \centering
    \begin{tabular}{llllr}
    \hline
        \textbf{(n,m)} & \textbf{Estimates} & \textbf{Avg Bias} & \textbf{MSEs} & \textbf{95\% CI}\\
         \hline
        \rule{0pt}{10pt}\multirow{2}{*}{(10,10)} & $\hat{\lambda_1}$=1.07666 & 0.07666 & 0.07914 &(-0.10393, 0.60624)\\
            & $\hat{\lambda_2}$=0.52771 & 0.02771 & 0.01282  \\
            \hline
     \rule{0pt}{10pt}\multirow{2}{*}{(15,15)} & $\hat{\lambda_1}$=1.03821 & 0.03821 & 0.03830 &(-0.12139, 0.63169)\\
            & $\hat{\lambda_2}$=0.51733 & 0.01753 & 0.00846 \\
            \hline
      \rule{0pt}{10pt}\multirow{2}{*}{(25,25)} & $\hat{\lambda_1}$=1.02654 & 0.02654 & 0.02262 &(-0.11027, 0.62711)\\
            & $\hat{\lambda_2}$=0.51750 & 0.01750 & 0.00496 \\
            \hline
       \rule{0pt}{10pt}\multirow{2}{*}{(30,30)} & $\hat{\lambda_1}$=1.02066 & 0.02066 & 0.01738 &(-0.02649, 0.53561)\\
            & $\hat{\lambda_2}$=0.50841 & 0.00841 & 0.00363 \\
            \hline
       \rule{0pt}{10pt}\multirow{2}{*}{(50,50)} & $\hat{\lambda_1}$=1.01033 & 0.01033 & 0.00939 &(0.07393, 0.43790)\\
            & $\hat{\lambda_2}$=0.50595 & 0.00595 & 0.00217 \\
            \hline
      \rule{0pt}{10pt}\multirow{2}{*}{(75,75)} & $\hat{\lambda_1}$=1.01027 & 0.01027 & 0.00697 &(0.09505, 0.41531)\\
            & $\hat{\lambda_2}$=0.50472 & 0.00472 & 0.00132 \\
            \hline
    \end{tabular}
\end{table}

\section{Applications}
In the following section, we examine two real data sets that show the breaking strength of jute fiber at different gauge lengths of 10mm and 20 mm (see Xia et al. (2009)). We present two datasets as follows:
\subsection*{\textbf{Dataset 1 : Jute fiber breaking strength of 10 mm}}
\noindent 693.73, 704.66, 323.83, 778.17, 123.06, 637.66, 383.43, 151.48, 108.94, 50.16,\\ 671.49, 183.16, 257.44, 727.23, 291.27, 101.15, 376.42, 163.40, 141.38, 700.74,\\ 262.90, 353.24, 422.11, 43.93, 590.48, 212.13, 303.90, 506.60, 530.55, 177.25.
\subsection*{\textbf{Dataset 2 : Jute fiber breaking strength of 20 mm}}
\noindent 71.46, 419.02, 284.64, 585.57, 456.60, 113.85, 187.85, 688.16, 662.66, 45.58,\\ 578.62, 756.70, 594.29, 166.49, 99.72, 707.36, 765.14, 187.13, 145.96, 350.70,\\ 547.44, 116.99, 375.81, 581.60, 119.86, 48.01, 200.16, 36.75, 244.53, 83.55.

Based on the assumption that the two independent samples have been drawn from $EGD(3, \lambda_1)$ and $EGD(3, \lambda_2)$, respectively, we provide the MLE estimates for parameters $\lambda_1$ and $\lambda_2$ and the goodness-of-fit tests Kolmogorov-Smirnov (K-S) and Cramer-Von Mises (CVM) in Table \ref{egdssfindings}. As a result, the MLE of R is $\hat{R}=0.5319$, and the 95\% CI for R is (0.3936,0.6702).

\begin{table}[!ht]
    \centering
    \caption{\textbf{MLE, CVM, and KS goodness of fit tests}}\label{egdssfindings}
    \begin{tabular}{lllr}
    \hline
        \textbf{Data} & \textbf{Estimates} & \textbf{CVM (p-value)} & \textbf{KS (p-value)} \\
         \hline
        \textbf{Length 10 mm} & 0.008149069 & 0.13151 (0.4533) & 0.1393 (0.5584)  \\ \hline
        \textbf{Length 20 mm} & 0.008725855 & 0.41935 (0.06361) & 0.20661 (0.1336)  \\ \hline
    \end{tabular}
\end{table}

\section{Conclusion}
There have been several well-developed estimation techniques for SS models with single components that follow well-known lifetime distributions. However, there has been no consideration of the EGD model. In this paper, we consider the problem of estimating SS reliability with an EGD distribution in a single-component SS model for independent stress and strength random variables. We obtain the MLE of SS reliability, $\hat{R}_{ML}$. Extensive simulation causes MSE and average biases caused by the estimates to approach zero when sample sizes are increased. Analyses are conducted on real-life datasets.

%\bibliography{mybibfile}

\begin{thebibliography}{}

\bibitem{alght2013}
\textrm{Al-Mutairi, D. K., Ghitany, M. E., and Kundu, D. (2013). Inferences on stress-strength reliability from Lindley distributions.} \emph{\textrm{Communications in Statistics-Theory and Methods 42(8):1443–1463.}}

\bibitem{Ballai2009}
\textrm{Balakrishnan, N. and Lai, C. D. (2009)}. \textrm{Continuous Bivariate Distributions, 2 edn, Springer, New York.}

\bibitem{birn1956}
\textrm{Birnbaum, Z.W. (1956). On a use of the Mann-Whitney statistic. In:}\emph{\textrm{ Proceedings of Third Berkeley Symposium on Mathematical Statistics and Probability}}, \textrm{vol. 1, pp. 13–17, University of California Press, Berkeley, CA.}

\bibitem{birMc1958}
\textrm{Birnbaum, Z.W., and McCarty, R.C. (1958). A distribution-free upper confidence bound for $P\{Y < X\}$, based on independent samples of X and Y}. \emph{\textrm{Annals of Mathematical Statistics, 29, 558–562.}}

\bibitem{chuhar1970}
\textrm{Church, J. D., and Harris, B. (1970). The estimation of reliability from stress-strength relationships.}\emph{\textrm{Technometrics 12, 49–54.}}

\bibitem{deepchakss2020}
\textrm{Deepthi, K. S., and Chacko, V. M. (2020)}. \textrm{Reliability estimation of stress-strength model using three parameter generalized Lindley distribution}.\emph{\textrm{ Advances and Applications in Statistics, 65(1), 69-89}}.

\bibitem{dey2017}
\textrm{Dey, S., Raheem, E., and Mukherjee, S. (2017)}. \textrm{Statistical properties and different methods of estimation of transmuted rayleigh distribution}. \emph{\textrm{Revista Colombiana de Estadística, 40(1), 165}}.

\bibitem{ghit2015}
\textrm{Ghitany, M. E., Al-Mutairi, D. K., and Aboukhamseen, S. M. (2015). Estimation of the Reliability of a Stress-Strength System from Power Lindley Distributions.}\emph{\textrm{ Communications in Statistics- Simulation and Computation, 44:1, 118-136.}}

\bibitem{gupbr2001}
\textrm{Gupta, R. C., and Brown, N. (2001). Reliability studies of the skew-normal distribution and its application to strength-stress models}. \emph{\textrm{Communications in Statistics-Theory and Methods, 30, 2427–2445.}}

\bibitem{hanagal2003}
\textrm{Hanagal, D., 2003}. \textrm{Estimation of system reliability in multicomponent series stress-strength models}. \emph{\textrm{Journal of the Indian Statistical Association, 41 (1), 1–7}}.

\bibitem{kot2003}
\textrm{Kotz, S., Lumelskii, Y., and Pensky, M. (2003). The stress-strength model and its generalizations: Theory and applications. Singapore: World Scientific Press.}

\bibitem{krish2007}
\textrm{Krishnamoorthy, K., Mukherjee, S., and Guo, H. (2007). Inference on reliability in two-parameter exponential stress-strength model}. \emph{\textrm{ Metrika 65:261–273.}}

\bibitem{kungup2005}
\textrm{Kundu, D. and Gupta, R.D. (2005). Estimation of $P[Y < X]$ for generalized exponential distribution}. \emph{\textrm{ Metrika, 61, 291-308.}}

\bibitem{kungup2006}
\textrm{Kundu, D. and Gupta, R.D. (2006), Estimation of $R = P[Y < X]$ for Weibull distributions}. \emph{\textrm{IEEE Transactions on Reliability, 55, 270-280.}}

\bibitem{kuRaq2009}
\textrm{Kundu, D. and Raqab, M.Z. (2009). Estimation of $R = P(Y < X)$ for three-parameter Weibull Distribution}. \emph{\textrm{ Statistics and Probability Letters, 79, 1839-1846.}}

\bibitem{Nadar2014}
\textrm{Nadar, M., Kizilaslan, F., and Papadopoulos, A., (2014)}. \textrm{Classical and Bayesian estimation of $P(Y<X)$ for Kumaraswamy’s distribution}. \emph{\textrm{Journal of Statistical Computation and Simulation, 84(7), 1505–1529}}.

\bibitem{raqk2005}
\textrm{Raqab, M. Z., and Kundu, D. (2005). Comparison of different estimators of $P[Y <X]$ for a scaled Burr type X distribution}. \emph{\textrm{ Communications in Statistics-Simulation and Computation 34, 465–483.}}

\bibitem{ramaku2008}
\textrm{Raqab, M. Z., Madi, M. D., and Kundu, D. (2008). Estimation of $P(Y < X)$ for the 3-parameter generalized exponential distribution.}\emph{\textrm{Communications in Statistics-Theory and Methods 37, 2854–2864.}}

\bibitem{sharma2015}
\textrm{Sharma, V.K., Singh, S.K., Singh, U., and Agiwal, V. (2015).} \textrm{The inverse lindley distribution: a stress-strength reliability model with application to head and
neck cancer data}. \emph{\textrm{Journal of Industrial and Production Engineering, 32(3), 162-173.}}

\bibitem{thomas2020}
\textrm{Thomas, B. and Chacko, V. M. (2020)}. \textrm{Exponential-Gamma$(3,\theta)$ Distribution and its Applications}. \emph{\textrm{Reliability: Theory and Applications, 15, 3(58), 49-61}}.

\bibitem{vc2022}
\textrm{Varghese, A. K. and Chacko V. M. (2022).} \textrm{Estimation of stress-strength reliability for Akash distribution}. \emph{\textrm{Reliability: Theory and Applications, 17, 3(69), 52-58}}.

\bibitem{xia2009}
\textrm{Xia, Z.P., Yu, J.Y., Cheng, L.D., Liu, L.F., and Wang, W.M. (2009)}. \textrm{Study on the breaking strength of jute fibers using modified Weibull distribution}. \emph{\textrm{Journal of Composites Part A: Applied Science and Manufacturing, 40, 54-59}}.

\end{thebibliography}
\end{document}